\documentclass{emulateapj}
\usepackage{apjfonts}

\shorttitle{First CMB Polarization Results from CAPMAP}
\shortauthors{Barkats et al.}

\journalinfo{{\sc The Astrophysical Journal, 619: L127-L130, 2005 February 1}}

\begin{document}

\submitted{} 
\title{First Measurements of the Polarization of the Cosmic Microwave Background Radiation at Small Angular Scales from CAPMAP}

\author{D.~Barkats\altaffilmark{1,2}, C. Bischoff\altaffilmark{3},
P.~Farese\altaffilmark{1,4}, L. Fitzpatrick\altaffilmark{3,5},
T.~Gaier\altaffilmark{6}, J.~O.~Gundersen\altaffilmark{7},
M.~M.~Hedman\altaffilmark{3,8,9}, L.~Hyatt\altaffilmark{1},
J.~J.~McMahon\altaffilmark{1}, D.~Samtleben\altaffilmark{3,9,10},
S.~T.~Staggs\altaffilmark{1}, K.~Vanderlinde\altaffilmark{3}, and
B.~Winstein\altaffilmark{3}}

\altaffiltext{1}{Department of Physics, Princeton University, Joseph Henry Laboratories, Jadwin Hall, P.O. Box 708, Princeton, NJ 08544-0708.}
\altaffiltext{2}{Current address: Department of Physics, California Institute of
Technology, 1200 East California Boulevard, MS 59-33, Pasadena, CA 91125.}
\altaffiltext{3}{Kavli Institute for Cosmological Physics and Enrico Fermi
Institute, University of Chicago, 5641 South Ingleside Avenue, Chicago, IL, 60637.}
\altaffiltext{4}{Dicke Fellow.}
\altaffiltext{5}{Current address: Department of Physics, Harvard University,
17A Oxford Street, Cambridge, MA 02138.}
\altaffiltext{6}{Jet Propulsion Laboratory, California Institute of
Technology, 4800 Oak Grove Drive, Pasadena, CA 91109.}
\altaffiltext{7}{Department of Physics, University of Miami, 
James L. Knight Physics Building, 1320 Campo Sano Drive, Coral Gables, 
FL 33146.}
\altaffiltext{8}{Current address: Department of Astronomy, Cornell University, 
Space Sciences Building, Ithaca, NY 14853-6801.}
\altaffiltext{9}{Kavli Fellow.}
\altaffiltext{10}{Corresponding author: dorothea@kicp.uchicago.edu.}

\begin{abstract}
Polarization results from the Cosmic Anisotropy 
Polarization MAPper (CAPMAP) experiment are reported.  These
are based on 433 hr, after cuts, observing a 2 deg$^2$
patch around the north celestial pole with four 90 GHz
correlation polarimeters coupled to optics defining $4\arcmin$
beams. The {\it E}-mode flat band-power anisotropy within
$l=940^{+330}_{-300}$ is measured as 66$^{+53}_{-39}~\mu$K$^2$; the
95$\%$ confidence level upper limit for {\it B}-mode power within
$l=1050^{+590}_{-520}$ is measured as 38~$\mu$K$^2$.
\end{abstract}


\section{INTRODUCTION}
The cosmic microwave background (CMB) is arguably the most fruitful
source of cosmological information. Its spectrum has been measured with extraordinary
precision \citep{FIRAS} and its spatial anisotropy has been very well
characterized, with the {\it Wilkinson Microwave Anisotropy Probe} \citep[WMAP;][]{WMAP:map} providing the most
comprehensive results. There are two distinct patterns to the
polarization of the CMB, conventionally termed {\it E}- and {\it B}-modes. The
former have recently been detected at the
degree scale \citep{Kovac02,WMAP:TE}; they
arise from the same density perturbations that dominate the
temperature anisotropies. The latter, which can result from
the lensing of {\it E}-modes due to the intervening
matter distribution or from primordial gravity waves, 
are expected to be far smaller. Characterization
of the {\it E}-mode power spectrum is important for testing the
understanding of the origin of the CMB as well as for breaking
degeneracies in cosmological parameter determinations
\citep[e.g.,][]{turok:I,Hu}.

In this Letter we present results from the first season of CAPMAP (Cosmic Anisotropy Polarization MAPper), 
an effort to characterize the {\it E}-mode polarization at 
90~GHz, where foregrounds are expected to be small, and at 4\arcmin,
where the polarization is expected to be near its
maximum.  In $\S\S$ \ref{sec:inst} and \ref{sec:obs} we
describe the instrument, its characterization and calibration, and the
observations.  In subsequent sections we summarize the reduction of the data
($\S$ \ref{sec:red}), the analysis ($\S$ \ref{sec:an}), and the systematic
studies ($\S$ \ref{sec:sys}). We conclude this Letter with a
discussion of the results and their implications ($\S$ \ref{sec:disc}).
\section{INSTRUMENT}
\label{sec:inst}
The CAPMAP instrument is described in detail
in \citet{instrument}; here its most important features are given.
In the winter of 2003, four W-band
(84$-$100 GHz) polarimeters (A$-$D) were installed in the
focal plane of the 7~m Cassegrain antenna \citep{bellabs} at the
Lucent Technologies facility in Crawford Hill, New Jersey (W$74^\circ11\arcmin11\arcsec\!\!$,
N$40^\circ23\arcmin31\arcsec\!\!$). The horn and meniscus lens system attached to each
polarimeter underilluminated the telescope mirrors with an edge taper
of $-35$~dB on the primary, resulting in a final beam of
$3\arcmin\!\!.88$ FWHM ($\sigma_B=1\arcmin\!\!.65$) with an
ellipticity less than 3$\%$. The polarimeters were housed inside a
single cryostat, nominally arranged so that A (C) observed $0^\circ\!\!.25$
below (above) and B (D) $0^\circ\!\!.25$ left (right) of the optical axis of
the telescope.

Heterodyne analog correlation polarimeters that multiply
two orthogonal components of the incident electric field produced an
output proportional to one linear combination of the Stokes
parameters {\it Q} and {\it U}. The two components derive from an ortho-mode
transducer (OMT) oriented with one axis horizontal.  The polarimeter
output is thus proportional to the difference in power for field
components aligned $\pm45^\circ$ to the horizontal in the focal plane.
The first-stage low noise amplifiers (LNAs) in each arm are InP HEMT
monolithic microwave-integrated circuits developed by JPL and Northrop
Grumman Space Technology \citep{northrop}. A mechanical refrigerator
cools the horns, the OMT, and the LNAs to $\leq$40 K. One local
oscillator inside the cryostat supplies the polarimeters with the 82~GHz
to down-convert the 84$-$100~GHz signals to an intermediate
frequency (IF) of 2$-$18~GHz. The oscillator signal in one arm of each
polarimeter is phase-switched at 4~kHz, well above the $1/f$ knee of
the amplifiers.  A detector diode monitors the total power in each
arm.  The phase-switched IF radiation is split into three 4~GHz-wide
subbands (S0, S1, S2) prior to multiplication, yielding then 12 polarization and eight total power channels.

The polarimeter outputs are digitized (24~bits, 100~kHz)
with a $\Sigma\Delta$ analog-to-digital converter (ICS-610); digitally
demodulated data are recorded at 100~Hz. The 100~kHz samples, averaged over 40 phase switches, are recorded to produce
quadrature data sets, demodulated out of phase with the phase switch,
which contain no celestial signal.
\section{OBSERVATIONS AND CALIBRATION}
\label{sec:obs}
\nopagebreak
Between 2003 February 18 and April 6, 541 hr of CMB data were
recorded. Calibration data were taken before, regularly throughout,
and after the season.

For the CMB observation the telescope's
optical axis moved through the north celestial pole (NCP) with an 8~s period by $\pm0^\circ\!\!.53$ at constant elevation. Polarimeter
offsets from the optical axis are given in Table \ref{tab:horns}; the
sampled region is a circle of diameter $1^\circ\!\!.6$
centered on the NCP, with the inner circle of diameter $0^\circ\!\!.14$
missing. This scan strategy yields quite uniform {\it Q/U} coverage over the
region.
\begin{table}[h]
\caption{Polarimeter Position Offsets$^a$}
\begin{center}
\begin{tabular}{lllll}
\hline & A & B & C & D
\\ \hline \hline co-elevation & -0.022 & -0.247 & -0.025 & 0.200
\\ cross-elevation  & -0.286 & -0.066 & 0.162 & -0.059
\end{tabular}
\end{center}
\label{tab:horns}
\footnotesize
$^a$ The polarimeter offsets from the optical axis are given in degrees. For the CMB observations, the constant-elevation scan was centered with the optical axis at $(-0^\circ\!\!.03, -0^\circ\!\!.09)$ from the NCP.
\end{table}

The pointing solution, determined with Jupiter observations, was
refined to $\pm30\arcsec$ using 150 observations of
astrophysical sources. The relative offsets, beam sizes and
the total power calibration were determined from the observations of Jupiter
in the total power channels.

Noise temperatures were estimated from several elevation scans
($20^\circ-90^\circ$). They ranged between 60 and 70~K and agreed
with laboratory cold load tests.

For the determination of polarized gains, a nutating plate was
installed in front of the secondary mirror to produce a modulated
polarized signal; this gave calibrations with an overall uncertainty
of 10$\%$ and a 3$\%$ relative uncertainty among the
channels. Frequent scans of the polarized source Tau A near
parallactic angle 110$^\circ$ (where the signal in the receivers is
maximal) provided a good calibration check
\citep{instrument}. Polarimeter sensitivities are summarized in Table
\ref{tab:sens}.

\begin{table}[h]
\begin{center}
\caption{Polarimeter Characteristics}
\label{tab:sens}
\begin{tabular}{rccccc}\hline
& \multicolumn{4}{c}{Polarimeter$^a$} & $\nu_{eff}^b$\\
Channel  & A & B & C & D & \\ \hline \hline
S0 & 3.2 & 3.4 & 2.9 & 3.1 & 87 \\
S1 & 2.1 & 2.1 & 2.1 & 2.2 & 92 \\
S2 & 2.7 & 2.0 & 3.5 & 2.5 & 96 \\ \hline
\end{tabular}
\end{center}
\footnotesize
$^a$ Sensitivities in mK$\sqrt{s}$ in thermodynamic units
\\ $^b$ Typical central frequency in GHz
\end{table}
The calibration was corrected for 
slow few kelvin variations in the IF electronics and
for a varying phase between the digitization and phase
switch clocks. 
The IF gains changed by up to 3$\%$~K$^{-1}$ but the temperature was stable to 1 K for 70$\%$ of the
data. The clock correction ranged from 0.9 to 1 and varied
by at most 0.003~hr$^{-1}$.  The calibration for two of the channels (DS0, DS2) was adjusted by $\sim10\%$ to compensate for gain
compression.
\section{DATA REDUCTION} 
\label{sec:red}
The CMB data were divided into 40~s periods; data within
each were divided into 20 $\sim3\arcmin$-wide, nearly uniformly populated azimuthal
bins, discarding the $9\%$ of the data at the turnarounds. Means and standard errors calculated in each
bin were carried forward in the analysis. Varying the period
between 16 and 40~s did not alter the final results.

A second-order polynomial was fitted to the 
azimuthal samples in each 40~s period and removed. The polarimetry channels reject the
common mode intensity by more than 23~dB, but their residual
sensitivity to the atmospheric temperature, which varied daily by 10~K, dominated the intercept terms. The linear terms were evident when
co-added over an hour, revealing slopes as large
as $500~\mu$K~deg$^{-1}$, with variations up to $200~\mu$K~deg$^{-1}$day$^{-1}$.
The quadratic terms were even smaller and more stable.

Abnormally high
fluctuations ($>4\sigma$) in the polarized channels or unusual changes
in the ratios between different total power channels were observed for 20 hr of data.
Both symptoms result from ice formation on the cryostat
windows so these data were removed. The five most sensitive total power channels were used to identify bad weather with a cut on their fractional deviations.  Only periods that passed the
cut for all five channels were retained, eliminating 88 hr and
leaving 433 hr for the final data sample.
Ten percent variations in this cut level changed the result within statistics.

Collapsing all polarization data into the
20 azimuthal bins reveals scan synchronous structures
(SSSs) with rms levels from 10 to 26~$\mu$K; from a 90~GHz side lobe beam map, these are likely due to
ground pickup. Dependences on variables like system temperature, local sidereal
time (LST), or ambient temperature were studied. Only BS1 and BS2 showed such 
a dependence, varying with the ambient temperature.
The mean of the data in each azimuthal bin was
removed, and residual effects that could arise from
correlations to ambient temperature are modeled as described below.
The channel with the largest structure and variation (BS2) is dropped
with negligible change in the results but with a significant reduction
in the systematic uncertainty.

\begin{figure}[t]
\plotone{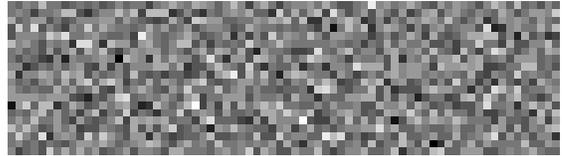}
\caption{The 20$\times$72 data vector for receiver C. The scale
ranges between $\pm 210$~$\mu$K; the average unchopped error for each
element is 64~$\mu$K. These pixels
are spaced approximately 0.8 beamwidths in azimuth; in the LST
direction, their spacing ranges from about 0.1 to 1 beam width
depending on the azimuthal position.}
\label{fig:map}
\end{figure}

After structure removal, corrections for atmospheric
absorption (typically 20$\%$) derived from the total power
channels, were applied. The corrected periods,
already parsed into azimuth bins, were binned into 72 LST bins, yielding 72$\times$20=1440 element data vectors for each of the
12 channels. 
The weighted averages
of the three frequency channels (which have correlations less than
1$\%$) were used in the likelihood analysis. One resulting data vector
is displayed in Figure \ref{fig:map}.
\section{DATA ANALYSIS}
\label{sec:an}
\begin{figure}[t]
\plotone{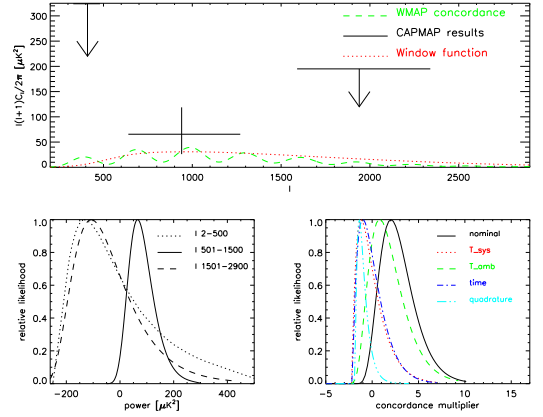}
\caption{Likelihood results. {\it Top}: Results from the three-band analysis, the concordance model ({\it long dashed line}), and the trace of the window function $W_l^E$, as
defined in equation (\ref{eq:win}), for receiver C, in arbitrary units (dotted line). For the outer bands, 95$\%$ confidence level upper limits are
displayed. {\it Lower left}:
Individual marginalized likelihood curves for the three bands. 
{\it Lower right}: Concordance model fitted to the data and
several null maps as described in the text.}
\label{fig:result}
\end{figure}

A Bayesian likelihood analysis was performed by maximizing the
likelihood, 
\begin{equation} \mathcal{L} = { 1\over \sqrt{\det(C)}}
\exp \{ -{\bf x}^T C^{-1} {\bf x}/2 \},
\end{equation}
where ${\bf x}$ is the data vector with $4 \times 1440$ entries and $C$ is the
covariance matrix. The vector ${\bf x}$ sums the signal ($s$) and
receiver noise ($n$) terms: $x_i=A_{ij}s_j+n_i$, where $A_{ij}$
determines the linear combination of {\it Q} ($s_1$) and {\it U} ($s_2$) in pixel
$i$.  Similarly, the covariance matrix comprises signal ($C_T$) added
to noise $N$: $C_{ij} = <x_i x_j> = C_{T,ij} + N_{ij}$, where $N_{ij}=n_i^2\delta_{ij}$, with typical noise
$n_i\approx 60\,\mu$K. The signal covariance depends on the multipole moments for {\it E}- and {\it B}-modes, $C_l^E$ and
$C_l^B$:
\begin{equation}
C_{T,ij} = \sum_l \frac{2l+1}{4\pi} (C_l^E W_{l,ij}^E + C_l^B W_{l,ij}^B ).
\label{eq:win}
\end{equation}
Here, the window function matrices $W_l^X$ account for (1) beam
functions $B_l^2\approx \exp(-l^2\sigma_B^2$), (2) the different
pixel areas in the data vector, and (3) the conversion from ({\it Q, U}) to $X=(E, B)$
\citep{ZS}. Figure \ref{fig:result} shows the trace of
receiver C's {\it E}-mode window function, similar to
those of the other receivers. The likelihood is maximized with respect
to one to three variables that parameterize $C^E_l$ and/or $C^B_l$,
as described below.

The signal-to-noise ratio eigenmode method was used \citep[e.g.,][]{bond98}: transformations into the basis where the noise matrix is
white and then into the eigenbasis of the
transformed $C_T$ are performed. Removing offsets every 40~s and
subtracting SSSs transform ${\bf x}$ to $D {\bf x}$. Accordingly, $C$ is
transformed to $DCD^T$, and the analysis is continued in the subspace
onto which $D$ projects.

The code was exercised extensively with simulated data
sets. Realizations with noise levels from that of the data to 20 times
smaller were superposed on CMB realizations of the concordance model
\citep{WMAP:parameters}. The likelihood estimator was found to be
unbiased; however, at the noise level of the data, the loss of
sensitivity from the offset removals is a factor of 2.

Four likelihood evaluations were made: first by taking $C^E_l$ as a
multiplier of the concordance model, with $C^B_l \equiv 0$; second
by taking $l(l+1)C^E_l$ to be a constant (flat band-power),
with $C^B_l \equiv 0$; third by simultaneously taking $C^E_l$ as
a multiplier to the concordance model and $l(l+1)C^B_l$ as a constant;
and fourth by taking $C^E_l$ to comprise three flat band-power levels,
in the $l$ bins (2-500/501-1500/1501-2900), with $C^B_l \equiv
0$ (three-band analysis). Fit results are summarized in Table
\ref{tab:result1}. Effective $l$s are calculated according to
\citet{Steinhardt} and quoted with the central 68$\%$ region of
$W_l/l$.

The various likelihood evaluations yield a 2~$\sigma$ detection of
the {\it E}-mode signal, with no {\it B}-mode signal or significant
dependence on frequency observed. Results from the
three-band analysis are shown in Figure
\ref{fig:result}. The upper band has negligible correlation with
the other two. For the lower two bands the marginalized results
from integrating over their likelihood contours are displayed. The
marginalization shifted the peak of the middle band down by 6$\%$,
increasing its width by 12$\%$. 
A fit of the middle $l$-band's likelihood to an offset
lognormal distribution \citep{bond} gives 
66~$\mu$K$^2$ with a variance of (44~$\mu$K$^2$)$^2$ and a noise related
offset of 93~$\mu$K$^2$.
\section{EVALUATION OF SYSTEMATIC UNCERTAINTIES}
\label{sec:sys}
The three most important sources of systematic uncertainty were
associated with the telescope pointing, the scan synchronous residual
structure and the relative gains. Their impact was
evaluated on fits of {\it E}-mode power to a multiplier of the concordance
model; selected studies confirmed that uncertainties estimated with
those fits apply to the three other fitting methods.

The effect of possible mispointing was determined by varying the
global pointing solution within its uncertainty, $\sim10\%$ of a beamwidth. 
The $1~\sigma$ shifts in azimuth and elevation were added in
quadrature, resulting in an uncertainty of $12\%$.

Effects of the SSSs were simulated. It was determined separately in each channel for two
halves of the data, divided according to ambient temperature and its extracted (linear) variation.
Multiple simulations were made with structure added to
the CMB according to the ambient temperature in each period.  This shifted the result upward by on average
0.1 (times concordance), and the spread, as the underlying CMB was
varied, was $\pm$0.1. The uncertainty is then conservatively taken to
be $\pm0.2$ (corresponding to a 10$\%$ uncertainty).

The effect of relative gain uncertainties was estimated from the
scatter in the result with the gain of each of the 11 polarimeters 
chosen randomly within the estimated $\pm 3$\% uncertainty, leading
to an uncertainty of $6\%$.

Three other sources of uncertainty were studied. The beam size was
uncertain to 2$\%$; varying it by even 10$\%$ leads to less than a 3$\%$
change in the result. As mentioned earlier, the polarimeters have a
weak response to temperature anisotropies ($-23$~dB), and the optics
produces a small quadrupole response at scales smaller than the beam
($-12$~dB).  Simulations show that these also lead to negligible effects
\citep[see also][]{HHZ}.

Several checks of the robustness of the result were performed. Excluding single radiometers or 25$\%$ 
data samples gave consistent results. Difference maps were created by
splitting the data into two parts according
to high and low system and ambient temperatures and time. Data vectors
were produced for each half (D$_1$ and D$_2$) and the likelihood was
evaluated on (D$_1-$D$_2$)/2. These likelihoods, together with that
for the quadrature data, are shown in Figure \ref{fig:result}; none
show a signal. Finally, a parallel analysis was
performed. Offsets were removed on a 4~s timescale, and an
independent code implemented a different pixelization
scheme and a different means of evaluating the associated covariance matrix and nulling
out the appropriate modes. It was used to obtain a multiplier to concordance
and gave a consistent result.

Adding the three main uncertainties in quadrature, the overall
systematic uncertainty becomes 17$\%$. The 10$\%$ overall gain
uncertainty contributes an additional 20$\%$ without reducing the
significance of the detection.
\begin{table}[h]
\begin{center}
\caption{CAPMAP results$^a$}
\label{tab:result1}
\begin{tabular}{lccccc}\hline \hline
\multicolumn{6}{c}{{\it E}-mode multiplier to concordance, $l$=1120$^{+560}_{-520}$, $C_l^B\equiv$0}
\\ \hline frequency & all$^a$ & S0 & S1 & S2 
\\ result$^b$ & 2.0$^{+2.1}_{-1.5}$ & 3.9$^{+7.3}_{-4.8}$ & -0.1$^{+2.8}_{-1.8}$ & 9.4$^{+8.8}_{-6.1}$ 
\\ \hline \multicolumn{6}{c}{Fits to flat band-power, $C_l^B\equiv$0 except for fit for {\it B}} 
\\ \hline & B$^c$ & E & E1$^d$ & E2$^d$ & E3$^d$
\\ $l$ & 1050$^{+590}_{-520}$ & 1120$^{+560}_{-520}$ & 410$^{+70}_{-80}$ & 940$^{+330}_{-300}$ & 1940$^{+400}_{-350}$
\\  result$^a$& $<38\mu$K$^2$ & 31$^{+39}_{-27}\mu$K$^2$ & $<324\mu$K$^2$ & 66$^{+53}_{-39}\mu$K$^2$ & $<195\mu$K$^2$
\end{tabular}
\end{center}
\footnotesize
$^{a}$ Simultaneous fit with {\it B} also gives 2.0$^{+2.1}_{-1.5}$
\\ $^b$ With 68$\%$ interval of highest posterior density or 95$\%$ upper limit
\\ $^{c}$ Fitted simultaneously with {\it E}
\\ $^{d}$ Three-band analysis
\end{table}
\section{DISCUSSION}
\label{sec:disc}
Evidence of polarization anisotropy in the CMB in the multipole
region around $l=1000$, where the {\it E}-mode power spectrum is expected to
peak, has been presented.  Signals of increasing significance are
seen when fitting for (1) a flat band-power, (2) a multiplier of the
expected concordance power spectrum, and (3) power in a band centered
at these high $l$-values.  The {\it E}-mode signal does not change when
fitting simultaneously for both {\it E}- and {\it B}-modes, and the latter are not
detected. Here the possible contaminations to these results are
discussed together with their significance.

Foregrounds are not expected to be significant. From
\citet{FDS} dust maps and {\it WMAP} synchrotron maps \citep{WMAP:foregrounds}
around the NCP, temperature anisotropies of 4.4
and 1.3~$\mu$K, respectively, are expected. Even with pessimistic
assumptions about polarization fractions and how these foregrounds
extrapolate to CAPMAP's angular scales, their contributions are well
below the observed signal.

There are no deep source maps exactly at the NCP.  A survey at
90 GHz \citep{MMA} extrapolates 4400 sources brighter
than 100 mJy over the entire sky, implying a 20$\%$ chance for
one to be in the CAPMAP field. From the instrument response
(3~$\mu$K~mJy$^{-1}$), such a source would contribute at the 20~$\mu$K level in
1 pixel if it had the same high polarization as Tau A (7.5$\%$).
Effects on the polarization power spectra, along with those from weaker but more
copious sources, are negligible. Conclusions from a dedicated source survey
near the NCP \citep{Leitch00} are similar.

Adding the systematic uncertainty in quadrature to the result in the
middle l-band, a null result is excluded at just over the 2~$\sigma$
level. Including the overall calibration uncertainty, this result is
consistent with the concordance model within 72$\%$ of the highest posterior
density. This result at 90 GHz is consistent with the very recent
results from the Degree Angular Scale Interferometer and Cosmic Background Imager at 30 GHz \citep{Leitch04, Readhead}.

\acknowledgements We thank Tom Crawford, Wayne Hu, Norm Jarosik, Stephan Meyer,
Lyman Page, and David Spergel for many helpful discussions and
Michelle Yeh and Eugenia Stefanescu for help in the data collection.  We thank
Lucent Technologies for use of the 7~m telescope and Bob Wilson, Greg Wright and Tod Sizer for their assistance with it.
This work was supported by NSF grants PHY-9984440,
PHY-0099493, PHY-0355328, AST-0206241 and PHY-0114422 and by the Kavli Foundation.
Portions of this work were carried out at the Jet Propulsion 
Laboratory, California Institute of Technology, operating under a 
contract from the National Aeronautics and Space Administration.

\end{document}